\documentclass{PoS}

\usepackage{epsfig,epsf}
\usepackage{bm} 
\newcommand{\beq}{\begin{equation}}
\newcommand{\beql}[1]{\begin{equation}\label{#1}}
\newcommand{\eeq}{\end{equation}}
\def\bal#1\gal{\begin{eqnarray}#1\end{eqnarray}}
\newcommand{\ball}[1]{\bal\label{#1}}
\newcommand{\eq}[1]{(\ref{#1})}
\newcommand{\fig}[1]{Fig.~\ref{#1}}

\newcounter{topiccounter}
\renewcommand{\b}[1]{{\bm #1}} 

\newcommand{\as}{\alpha_s}

\newcommand{\im}{\,\mathrm{Im}\,}

\newcommand{\aver}[1]{\left\langle #1 \right\rangle}

\title{Coulomb corrections to DIS off heavy nuclei}

\ShortTitle{Coulomb corrections  to DIS off heavy nuclei}

\author{\speaker{Kirill Tuchin}\thanks{This work  was supported in part by the U.S. Department of Energy under Grant No.\ DE-FG02-87ER40371.}\\
       Department of Physics and Astronomy, Iowa State University, Ames, Iowa, 50011, USA\\
        E-mail: \email{tuchin@iastate.edu}}

\abstract{
An essential part of experimental program at the future  Electron Ion Collider is the study of the nuclear structure and dynamics at low $x$. DIS at low $x$ is characterized by large longitudinal coherence length that by far exceeds radii of heavy nuclei. The coherent behavior is essential feature of the nuclear matter at low $x$. This pertains not only to the strong interactions, but also to electromagnetic ones. Coherent interactions of a projectile with nucleons in a heavy nucleus are characterized by parameters $\as^2 A^{1/3}\sim 1$ and $\alpha Z\sim 1 $ in strong and electromagnetic interactions respectively. Contributions exhibiting non-trivial dependence on $\alpha Z$ are called the Coulomb corrections. We compute the Coulomb corrections to the cross sections of the semi-inclusive  and diffractive DIS. We show that they violate  the geometric scaling in a wide range of photon virtualities and is weakly $x$-independent. In heavy nuclei at low $Q^2$ the Coulomb correction to the total and diffractive cross sections is about 20\% and 40\% correspondingly. }

\FullConference{XXIII International Workshop on Deep-Inelastic Scattering,\\
		27 April - May 1 2015\\
		Dallas, Texas}

\begin{document}

\section{Why electromagnetic contribution is important}

DIS at low $x$ is characterized by large longitudinal coherence length that by far exceeds radii of heavy nuclei.  In QCD, color fields of nucleons in a heavy nucleus fuse to create an intense coherent color field which  has fundamental theoretical and phenomenological importance. Since nuclear force is short-range, only nucleons along the same impact parameter add up to form the coherent field. Because the QCD contribution to the scattering amplitude at high energy  is approximately imaginary, it is proportional to $\as^2$. Thus, the parameter that characterizes the color-coherent field is $\as^2 A^{1/3}\sim 1$, where $A$ is atomic weight. 

Along with strong color field, heavy-ions also posses strong electromagnetic Coulomb field. The electromagnetic force is long range, therefore all $Z$ protons of an ion contribute to the field. The QED contribution to the scattering amplitude is approximately real. As a result, the parameter that characterizes the coherent electromagnetic field is $\alpha Z\sim 1$. Since both parameters $\as^2 A^{1/3}$ and $\alpha Z$ are of the same order of magnitude in heavy ions,  electromagnetic force must be taken into account along with the color one. This observation is a direct consequence of coherence which enhances the electromagnetic contribution by a large factor $Z$. Such electromagnetic contributions are called the Coulomb corrections. 

In this article I focus on the Coulomb corrections to the total semi-inclusive and diffractive cross sections. At low $x$ the total $\gamma^*A$ cross section can be expressed in terms of the total dipole--nucleus cross section $\hat \sigma$ as follows (see e.g.\ \cite{Kovchegov:2012mbw})
\ball{a11}
\sigma_{T/L}(x,Q^2)= \frac{1}{4\pi}\int_0^1 dz \int d^2r\, \Phi_{T/L}(r,z)\,\hat \sigma(x,r)\,,
\gal  
where  $Q^2$ is the photon virtuality.  The light-cone wave functions for transverse and longitudinal polarizations of photon are given by 
\ball{a13}
\Phi_{T}= \sum_f\frac{2\alpha N_c}{\pi}\left\{ [z^2+(1-z)^2]a^2 K_1^2(ar) +m_f^2 K_0^2(ar)\right\}\,, &\\
\Phi_L= \sum_f\frac{2\alpha N_c}{\pi} \, 4Q^2z^2(1-z)^2K_0^2(ar)\,, &\label{a14}
\gal
where  $m_f$ is quark mass, $z$ is the fraction of the photon's light-cone momentum  carried by the quark, $r$ is the size of the $q\bar q$ dipole in the transverse plane and $a^2= z(1-z)Q^2+m_f^2$. \footnote{The relationship between  the cross section $\sigma=\sigma_T+\sigma_L$ and $F_1$, $F_2$ structure functions is non-trivial due to large Coulomb corrections to the leptonic tensor  \cite{Kopeliovich:2001dz}.}  The dipole cross section $\hat \sigma$ depends non-trivially on $A$ and $Z$. In the following sections I explain how it can be calculated. More details can be found in \cite{Tuchin:2013eya,Tuchin:2014fqa}.

\section{Nuclear dependence of the dipole cross section $\hat \sigma$. }

At high energies, interaction of the projectile proton with different nucleons  is independent inasmuch as the nucleons do not overlap in the longitudinal direction. This assumption is tantamount to taking into account only two-body interactions, while neglecting the many-body ones \cite{Glauber:1987bb}. In this approximation we can write the dipole-nucleus elastic scattering amplitude as 
\beql{b15}
\Gamma(\b b)=1- \exp\Big\{-\sum_a\langle \Gamma^{(1)}(\b b)\rangle\Big\}.
\eeq
where $\Gamma^{(1)}(\b b)$ is the dipole-nucleon elastic scattering amplitude and $\aver{\ldots}$ denotes average over the nucleon position in the nucleus.  For the sake of brevity, dependence of the scattering amplitudes on $x$ and $r$ is not explicitly indicated. The total dipole cross section can be computed using the optical theorem as follows 
\beql{b16}
\hat\sigma= 2\int d^2b \im[i\Gamma(\b b)]\,.
\eeq

Strong and electromagnetic contributions decouple in the elastic  scattering amplitude at the leading order in respective couplings:
\beql{b17} 
\Gamma^{(1)}= \Gamma^{(1)}_{s}+\Gamma^{(1)}_{em}\,.
\eeq
This is because  $i\Gamma^{(1)}_{em}$ is real, while $i\Gamma^{(1)}_{s}$ is imaginary, as discuss below. Owing to \eq{b17} we can cast \eq{b15} in the form  
\beql{b19}
\Gamma(\b b)=1- \exp\Big\{- A \aver{\Gamma^{(1)}_{s}(\b b)} -Z\aver{\Gamma^{(1)}_{em}(\b b)  }\Big\}\,.
\eeq
Averaging over the nucleus wave function is done as follows
\beql{b22}
\aver{\Gamma^{(1)}_{s}(\b b)}= \frac{1}{A}\int_{-\infty}^\infty dz_a \int d^2b_a\, \rho(\b b_a,z_a)  \Gamma^{(1)}_{s}(\b b-\b b_a)\,,
\eeq 
where  $\rho$ is the nuclear density.  Neglecting the diffusion region,  nuclear density is approximately constant  $\rho = A/(\frac{4}{3}\pi R_A^3)$ for points inside the nucleus and zero otherwise. The range of the nuclear force is about a fm, which is much smaller than the radius $R_A$ of a heavy nucleus. Therefore, $\b b\approx \b b_a$ and 
\beql{b25}
\aver{\Gamma^{(1)}_{s}(\b b)}= \frac{1}{A}\, 2\sqrt{R_A^2-b^2}\, \pi R_A^2\, \rho \Gamma^{(1)}_{s}(0)
=\frac{2C_F}{AN_c}\rho T(b)\frac{1}{2}\,\pi r^2\, \as^2\ln \frac{1}{r\mu}\,, 
\eeq
where $T(b)=2\sqrt{R_A^2-b^2}$ is the thickness function and $\mu$ is an infra-red scale. It follows from \eq{b25} that 
$A\langle\Gamma^{(1)}_{s}\rangle\sim \as^2 A^{1/3}$, which implies that \eq{b19} sums up  terms of order $\as^2 A^{1/3}\sim 1$ at $\as\ll 1$. Indeed, the leading strong-interaction contribution to the $\gamma^*N$ elastic scattering amplitude  corresponds to double-gluon exchange. Note also, that  the corresponding  $\langle i\Gamma^{(1)}_{s}\rangle$ is purely imaginary.  

To calculate the electromagnetic contribution to the $\gamma^*N$ scattering, note that the proton density in the nucleus is $Z\rho /A$. Hence
\bal\label{b29}
\aver{\Gamma^{(1)}_{em}(\b b)}&= \frac{1}{Z}\int_{-\infty}^\infty dz_a \int d^2b_a\, \frac{Z}{A}\rho(\b b_a,z_a)  \Gamma^{(1)}_{em}(\b b-\b b_a)\\
&=
 \frac{1}{iA}\rho\, 2 \alpha  \int d^2b_a\, T(b_a)\,\ln 
\frac{|\b b-\b b_a-\b r/2|}{|\b b-\b b_a+\b r/2|}\,.\label{b30}
\gal
The leading electromagnetic contribution to elastic  $\gamma^*N$ scattering amplitude arises from one photon exchange; the corresponding $\langle i\Gamma^{(1)}_{em}\rangle$ is purely real. We note, that \eq{b22} sums up terms of order $\alpha Z\sim 1$ at $\alpha\ll 1$. Had we been interested in purely electromagnetic scattering (e.g.\ of $e^-e^+$ instead of $q\bar q$) we could have approximated  $b\gg b_a\sim R_A$ owing to the long-range nature of the Coulomb potential. That would have yielded the well-known Bethe-Heitler-Maximon result \cite{Bethe:1934za,Bethe:1954zz} as shown in \cite{Tuchin:2009sg}. However, in DIS $b\sim R_A$ and no such approximation is possible. It follows from \eq{b16} and \eq{b19} that 
\beql{b33}
\hat\sigma= 
2\int d^2b\left\{ 1-\exp[-A \aver{\Gamma^{(1)}_{s}(\b b)}]\cos[Z\aver{i\Gamma^{(1)}_{em}(\b b)}]\right\}\,.
\eeq
Integrals in \eq{b30} and \eq{b33} can be analytically calculated in a simple but quite  accurate ``cylindrical nucleus" model (see e.g.\ \cite{Kovchegov:2001sc,Kharzeev:2004yx}), which approximates   the nuclear thickness function by the step function, viz.\  $T(b)=2R_A$ if $b<R_A$ and  zero otherwise. The result is \cite{Tuchin:2013eya}
\ball{a21}
&\hat\sigma(x,r)=\hat\sigma_{s}(x,r)+ \hat\sigma_{em}(x,r)\,,\\
&\hat\sigma_{s}(x,r)= 2\pi R_A^2 \left\{ 1-\exp\left[-\frac{1}{4}\tilde Q_s^2(x) r^2\right]\right\},\label{a22}\\
&\hat\sigma_{em}(x,r)= 4\pi r^2 (\alpha Z)^2 \ln\frac{W^2}{4m_f^2m_NR_A},\label{a23}
\gal
where $m_N$ is nucleon mass, $W$ is the $\gamma^* A$  center-of-mass energy given by $W^2=Q^2/x+m_N^2$ and $\tilde Q_s^2$ is the \emph{quark} saturation momentum (see e.g.\ \cite{Kovchegov:2012mbw}). 

Logarithm  that appears in \eq{a23} is the result of integration over the impact parameter from $R_A$ up to a cutoff $b_{max}$, which delimits the region of validity of the Weizs\"acker-Williams approximation. It is given by $b_{max}=\max\{W^2z(1-z)/(m_N(m_f^2+\b k^2))\}$, where $\b k$ is the quark's transverse momentum \cite{Tuchin:2009sg}. The largest size of the $q\bar q$ dipole, corresponding to the smallest $\b k$, is  $\sim 1/m_f$  due to the confinement. For that reason  $b_{max}$, and hence \eq{a23},  depends on the constituent quark mass $m_f$ rather than on the much smaller current quark mass $m_q$.

\section{Evolution effects}

Eqs.~\eq{a21}--\eq{a23} are derived in the quasi-classical approximation where the quark saturation momentum $\tilde Q_s^2$, and hence the QCD contribution to the total cross section, is $x$-independent. At lower $x$, such that $\as \ln(1/x)\sim 1$, the QCD quantum evolution effects become important and are described by the BK equation \cite{Balitsky:1995ub,Kovchegov:1999yj}. According to the BK equation the saturation momentum acquires $x$-dependence in the form $\tilde Q_s^2\sim A^{1/3}x^{-\lambda}$, where $\lambda$ is a certain positive number \cite{Levin:2001cv}. The functional form of the dipole cross section is also evolving with $x$;  \eq{a22} in that case is the initial condition.  There are several phenomenological models that describe the evolved dipole cross section. In this article I use the Golec-Biernat--Wusthoff model \cite{GolecBiernat:1998js} which retains the functional form of \eq{a22} while models the saturation momentum as follows 
\ball{e11}
\tilde Q_s^2= Q_0^2\left(\frac{x_0}{x}\right)^\lambda\,,
\gal
where  $Q_0=1$~GeV, $x_0=3.04\cdot 10^{-4}$, $\lambda=0.288$,  $R_A= R_pA^{1/3}$ with  $R_p=3.1$~GeV$^{-1}$, $m_f=140$~MeV, and $N_f=3$.  If we neglect the electromagnetic term \eq{a23} and use \eq{a22} in \eq{a11}, then we immediately observe that the total $\gamma^*A$ cross section exhibits the geometric scaling at $Q^2\gg m_f^2$. This is because $x$-dependence arises only through the combination $r^2Q_s^2(x)$, and the dipole size $r$ is determined by $1/Q$. 

That the Coulomb correction violates the geometric scaling is evident from \eq{a23} which, being an electromagnetic contribution,  does not depend on the strength of the color field determined by $\tilde Q_s^2$.  Unlike the QCD term \eq{a22}, the QED one \eq{a23} does not evolve much with $x$. Indeed, at the leading-log oder $\Gamma_{s/em}\sim (1/x)^{1+\Delta_{s/em}}$, where  $\Delta_{s}=4\ln 2(\as N_c/\pi)$ \cite{Balitsky:1978ic,Kuraev:1977fs} and   $\Delta_{em}= (11/32) \pi \alpha^2$ \cite{Gribov:1970ik,Mueller:1988ju}. Because $\Delta_{em}\ll \Delta_{s}$ we can neglect the effect of the QED evolution. 

\section{Total cross section}\label{sec:a}

Substituting \eq{a23} into \eq{a11} and integrating over $r$ we obtain the following analytic expression for the Coulomb correction to the total $\gamma^*A$ cross section \cite{Tuchin:2014fqa}
\ball{b11}
&\sigma_{{em},T/L}= (\alpha Z)^2\ln\frac{W^2}{4m_f^2m_NR_A} \sum_f \frac{4\alpha N_c}{3m_f^2}\, g_{T/L}(Q/m_f)\,,\\
&g_T(\eta)=\left[ 4 \left(\eta ^4+7 \eta ^2+8\right) \tanh ^{-1}\left(\frac{\eta  \sqrt{\eta ^2+4}}{\eta ^2+2}\right)
-2 \eta  \sqrt{\eta ^2+4} \left(\eta
   ^2+8\right)\right]\left[\eta ^3 \left(\eta ^2+4\right)^{3/2}\right]^{-1},\label{b13}\\
&g_L(\eta)=4\left[ \eta  \sqrt{\eta ^2+4} \left(\eta ^2+6\right)  
+4 \left(\eta ^2+3\right) \ln \frac{\eta -\sqrt{\eta
   ^2+4}}{\eta+\sqrt{\eta ^2+4}}\right]  \left[ \eta ^3 \left(\eta ^2+4\right)^{3/2}\right]^{-1}.\label{b14}
\gal  
Coulomb correction to the total semi-inclusive cross section \eq{b11} does not scale with $\tilde Q_s^2/Q^2$ and therefore explicitly violate the geometric scaling.

\begin{figure}[ht]
     \includegraphics[width=8cm]{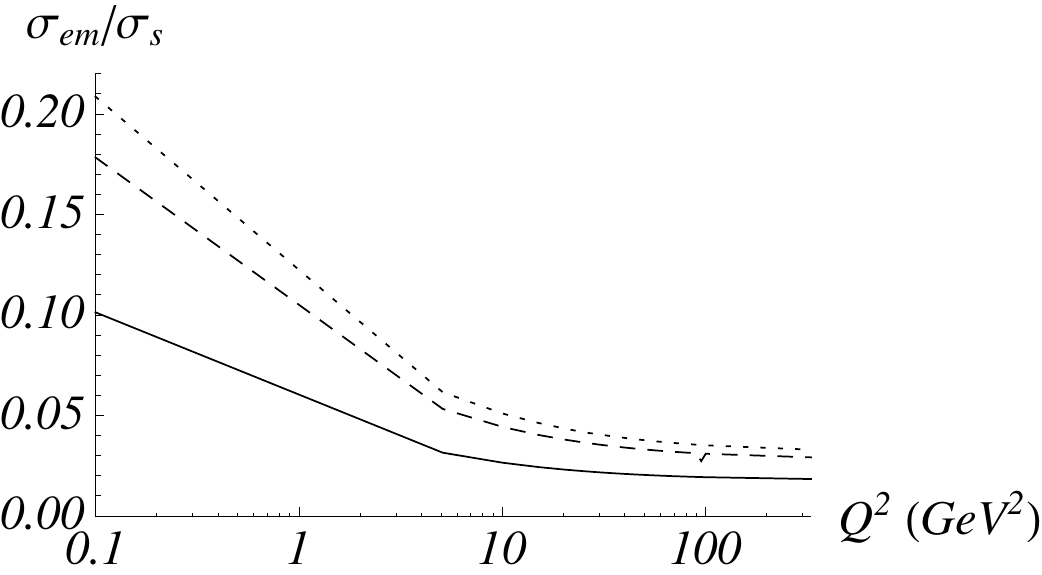}
 \caption{Ratio of QED and  QCD contributions to the total $\gamma^*A$ cross section at $x=10^{-4}$ as a function of $Q^2$ for silver (solid line), gold (dashed line) and uranium (dotted line) nuclei.  }
\label{fig1}
\end{figure}

\begin{figure}[ht]
     \includegraphics[width=8cm]{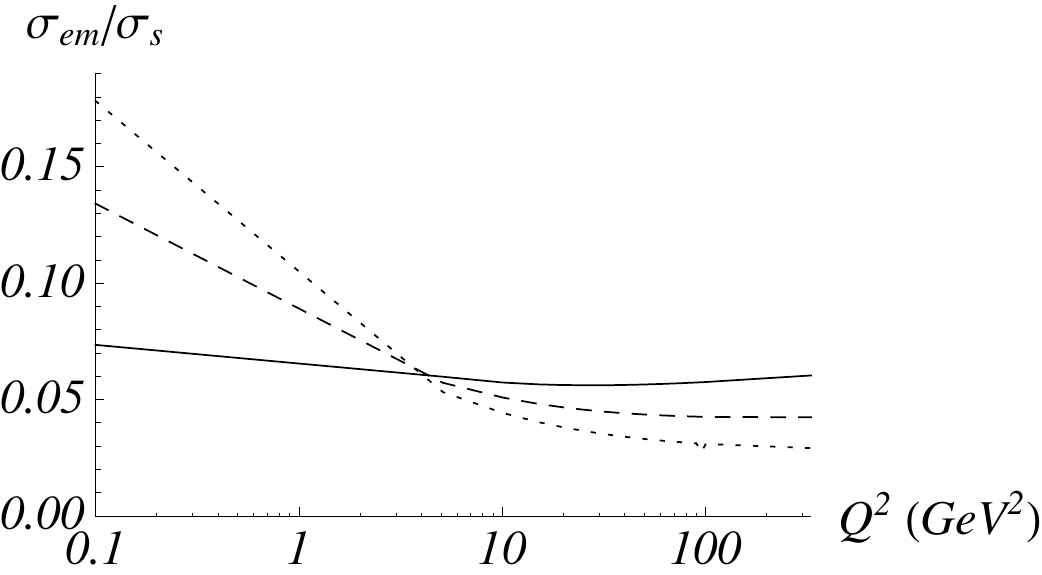} 
 \caption{Ratio of QED and  QCD contributions to the total $\gamma^*A$ cross section as a function of $Q^2$ for gold nucleus   at $x=10^{-2}$ (solid line), $x=10^{-3}$ (dashed line), $x= 10^{-4}$ (dotted line).}
\label{fig2}
\end{figure}

 The numerical results are shown in Figs.~\ref{fig1}--\ref{fig2}. We can see in \fig{fig1}  that at low $Q^2$ the QED correction for uranium nucleus at $x=10^{-4}$ can be as large as 20\% in the total cross section. It is remarkable that the Coulomb correction is non-negligible even at high $Q^2$. One should, however, interpret cautiously the results of our calculation at high $Q^2$ since  the model that we are using does not properly account for the DGLAP evolution. As expected, the relative size of Coulomb corrections increases with the nuclear weight and weakly depends on $x$.

\section{Diffractive cross section}\label{sec:d}

Total diffractive cross section corresponds to  elastic scattering of color dipole on the nucleus. It can be written as 
\ball{d11}
\sigma^{{diff}}_{T/L}(x,Q^2)= \frac{1}{4\pi}\int_0^1 dz \int d^2r\, \Phi_{T/L}(r,z)\,\hat \sigma^{el}(x,r)\,,
\gal  
where the total elastic dipole--nucleus cross section reads
\ball{d13}
&\hat \sigma^{el}(x,r)=\int d^2b \left| 1- \exp\left[- A \aver{i\Gamma_{s}} -Z\aver{i\Gamma_{em} }\right]\right|^2
\gal
A simple calculation yields \cite{Tuchin:2013eya}
\ball{d25}
\hat \sigma^{el}(x,r) =\hat \sigma_{s}^{el}(x,r)+ \hat \sigma_{em}(x,r)\,,
\gal
where $\hat \sigma_{em}$ is the QED contribution given by \eq{a23}, while  the QCD contribution is
\ball{d27}
\hat \sigma_{s}^{el}(x,r)=\pi R_A^2 \left\{ 1-\exp\left[-\frac{1}{4}\tilde Q_s^2(x) r^2\right]\right\}^2\,.
\gal

\begin{figure}[ht]
     \includegraphics[width=8cm]{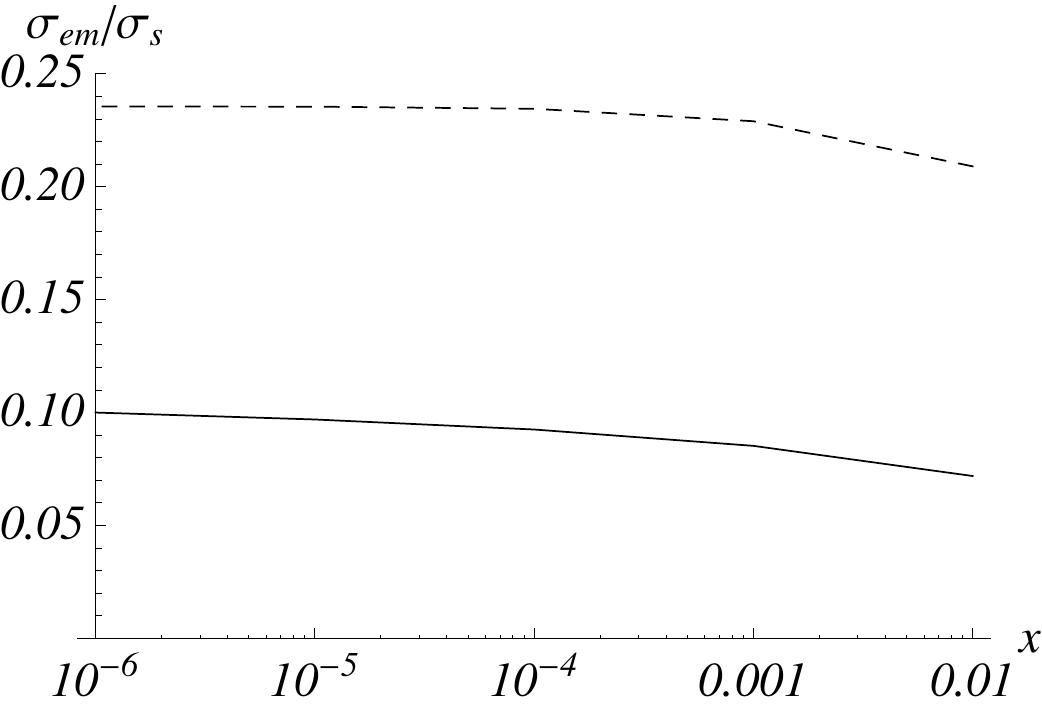} 
 \caption{Ratio of QED and  QCD contributions to the total (solid line) and diffractive (dashed) $\gamma^*A$ cross section as a function of $x$ for gold nucleus at  $Q^2=1$~GeV$^2$.}
\label{fig3}
\end{figure}

\begin{figure}[ht]
     \includegraphics[width=8cm]{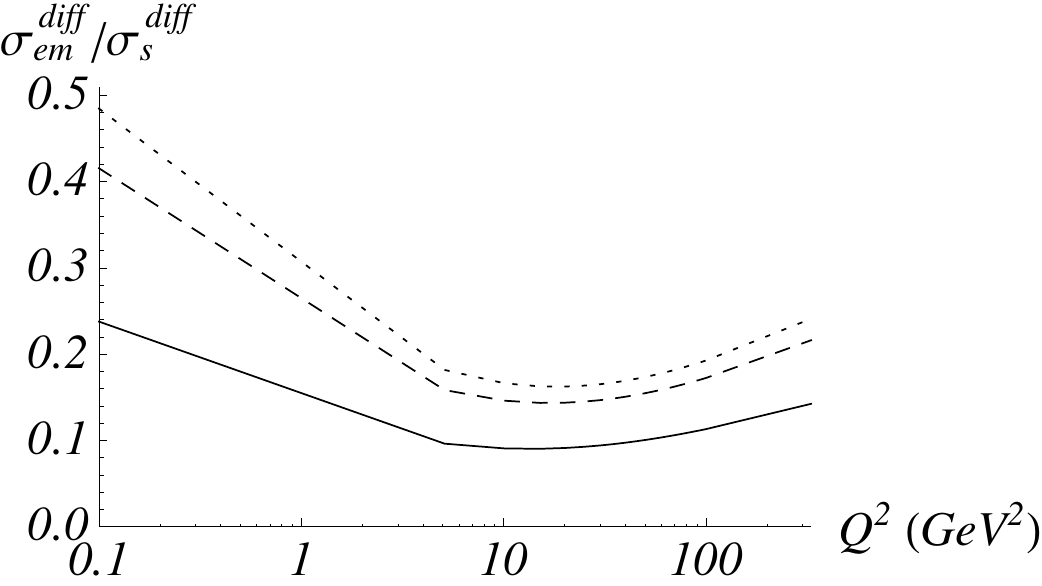}
 \caption{Ratio of QED and  QCD contributions to the diffractive $\gamma^*A$ cross section at $x=10^{-4}$ as a function of $Q^2$ for silver (solid line), gold (dashed line) and uranium (dotted line).  }
\label{fig4}
\end{figure}

Numerical results are shown in Figs.~\ref{fig3}--\ref{fig4}. We can see in \fig{fig4} that at low $Q^2$ the QED correction for uranium nucleus at $x=10^{-4}$ can be as large as 40\% in the diffractive cross section. The Coulomb corrections to the diffractive cross section are much larger than in semi-inclusive one, see \fig{fig3}. The reason is that at the leading order in coupling, electromagnetic interaction is elastic.

\section{Conclusion: beware of Coulomb corrections}
  
An essential part of experimental program at the future  Electron Ion Collider (EIC) is the study of the nuclear structure and dynamics at low $x$. Nuclear matter at low $x$ exhibits highly coherent behavior, the most striking manifestation of which is the  geometric scaling \cite{Stasto:2000er,Gribov:1984tu,Levin:1999mw,Levin:2000mv,Iancu:2002tr}. In semi-inclusive DIS the geometric scaling means that the total $\gamma^*p$ and $\gamma^*A$ cross sections scale with a dimensionless ratio $Q^2/\tilde Q_s^2(x)$, where $Q^2$ is photon virtuality and $\tilde Q_s(x)$ is the quark saturation momentum. While the geometric scaling is a signature of coherence in strong interactions, Coulomb corrections is a manifestation of coherence in electromagnetic interactions. Moreover, in DIS off heavy nuclei the two coherence effects are delicately entangled and separate only in a crude cylindrical nucleus model.  

Results presented in this work indicate that Coulomb corrections play an important role in the low $x$ DIS off heavy nuclei in a very wide range of $Q^2$ and $x$. In particular, they violate the geometric scaling. Since the focus of the EIC program is on the  strong interactions, one would like to learn how to subtract the Coulomb corrections from the measured quantities. We showed that Coulomb corrections can in principle be reliably computed. Since we used an oversimplified model for nuclear wave function that allowed us to derive analytical formulas, it is important to  refine our estimates by using more realistic nuclear profiles and more sophisticated low $x$ evolution models.



\begin{thebibliography}{99}

\bibitem{Kovchegov:2012mbw} 
  Y.~V.~Kovchegov and E.~Levin,
  ``Quantum chromodynamics at high energy,'' Cambridge University Press, 2013.
  
\bibitem{Tuchin:2013eya} 
  K.~Tuchin,
  Phys.\ Rev.\ C {\bf 89}, no. 2, 024904 (2014)
  [arXiv:1311.1124 [hep-ph]].


\bibitem{Tuchin:2014fqa} 
  K.~Tuchin,
  Phys.\ Rev.\ Lett.\  {\bf 112}, no. 7, 072001 (2014)
  [arXiv:1402.0771 [hep-ph]].

\bibitem{Kopeliovich:2001dz} 
  B.~Z.~Kopeliovich, A.~V.~Tarasov and O.~O.~Voskresenskaya,
  Eur.\ Phys.\ J.\ A {\bf 11}, 345 (2001)
  
\bibitem{Glauber:1987bb}
  R.~J.~Glauber,
{\it  In *Lo, S.Y. (ed.): Geometrical pictures in hadronic collisions*, 83-182. World Scientific} (1987).



\bibitem{Bethe:1934za}
  H.~Bethe and W.~Heitler,
  Proc.\ Roy.\ Soc.\ Lond.\  A {\bf 146}, 83 (1934).

\bibitem{Bethe:1954zz}
  H.~A.~Bethe and L.~C.~Maximon,
  Phys.\ Rev.\  {\bf 93}, 768 (1954).
  
  
\bibitem{Tuchin:2009sg} 
  K.~Tuchin,
  Phys.\ Rev.\ D {\bf 80}, 093006 (2009)
  [arXiv:0907.5189 [hep-ph]].

\bibitem{Kovchegov:2001sc} 
  Y.~V.~Kovchegov and K.~Tuchin,
  Phys.\ Rev.\ D {\bf 65}, 074026 (2002)
  
\bibitem{Kharzeev:2004yx} 
  D.~Kharzeev, Y.~V.~Kovchegov and K.~Tuchin,
  Phys.\ Lett.\ B {\bf 599}, 23 (2004)
  




\bibitem{Balitsky:1995ub}
  I.~Balitsky,
  Nucl.\ Phys.\  B {\bf 463}, 99 (1996)

\bibitem{Kovchegov:1999yj}
  Y.~V.~Kovchegov,
  Phys.\ Rev.\  D {\bf 60}, 034008 (1999)
  
\bibitem{Levin:2001cv}
  E.~Levin and K.~Tuchin,
  Nucl.\ Phys.\  A {\bf 693}, 787 (2001)
  
\bibitem{GolecBiernat:1998js} 
  K.~J.~Golec-Biernat and M.~Wusthoff,
  Phys.\ Rev.\ D {\bf 59}, 014017 (1998)


\bibitem{Balitsky:1978ic}
  I.~I.~Balitsky and L.~N.~Lipatov,
  Sov.\ J.\ Nucl.\ Phys.\  {\bf 28} (1978) 822
  [Yad.\ Fiz.\  {\bf 28} (1978) 1597].

\bibitem{Kuraev:1977fs}
  E.~A.~Kuraev, L.~N.~Lipatov, and V.~S.~Fadin,
  Sov.\ Phys.\ JETP {\bf 45}, 199 (1977)
  [Zh.\ Eksp.\ Teor.\ Fiz.\  {\bf 72}, 377 (1977)].


\bibitem{Gribov:1970ik} 
  V.~N.~Gribov, L.~N.~Lipatov and G.~V.~Frolov,
  Sov.\ J.\ Nucl.\ Phys.\  {\bf 12}, 543 (1971)
  [Yad.\ Fiz.\  {\bf 12}, 994 (1970)].
  
\bibitem{Mueller:1988ju}
  A.~H.~Mueller,
  Nucl.\ Phys.\  B {\bf 317}, 573 (1989).








\bibitem{Stasto:2000er} 
  A.~M.~Stasto, K.~J.~Golec-Biernat and J.~Kwiecinski,
  Phys.\ Rev.\ Lett.\  {\bf 86}, 596 (2001)
  
  
\bibitem{Gribov:1984tu} 
  L.~V.~Gribov, E.~M.~Levin and M.~G.~Ryskin,
  Phys.\ Rept.\  {\bf 100}, 1 (1983).


\bibitem{Levin:1999mw}
  E.~Levin and K.~Tuchin,
  Nucl.\ Phys.\  B {\bf 573}, 833 (2000)

\bibitem{Levin:2000mv}
  E.~Levin and K.~Tuchin,
  Nucl.\ Phys.\  A {\bf 691}, 779 (2001)


   
\bibitem{Iancu:2002tr}
  E.~Iancu, K.~Itakura and L.~McLerran,
  Nucl.\ Phys.\ A {\bf 708} (2002) 327




\end{thebibliography}
\end{document}